\documentclass{superfri}

\usepackage{subcaption}
\usepackage[utf8]{inputenc}

\usepackage{algorithm,algorithmicx}
\usepackage{algpseudocode}
\usepackage{listings}
\usepackage{courier}
\usepackage{tikz}
\usetikzlibrary{intersections}
\usetikzlibrary{calc}
\usepackage{tkz-euclide}
\usetkzobj{all}
\usepackage{pgfplots}
\usepackage{pgfplotstable}
\usepackage{xifthen}
\usepackage{scalerel}
\usepackage{multirow}
\usepackage[hidelinks]{hyperref}
\usepackage{booktabs}

\begin{document}

\author{Julian Hornich\footnote{\label{fau}Erlangen Regional Computing Center (RRZE) at Friedrich-Alexander-Universit\"at Erlangen-N\"urnberg (FAU), Germany}, Julian Hammer\footnoteref{fau}, Georg Hager\footnoteref{fau}, Thomas Gruber\footnoteref{fau}, Gerhard Wellein\footnote{\label{fdi}Department for Computer Science at  Friedrich-Alexander-Universit\"at Erlangen-N\"urnberg (FAU), Germany}}

\title{Collecting and Presenting Reproducible\\Intranode Stencil Performance: INSPECT}

\maketitle{}

\begin{abstract}
Stencil algorithms have been receiving considerable interest in HPC research for decades. The
techniques used to approach multi-core stencil performance modeling and engineering span basic
runtime measurements, elaborate performance models, detailed hardware counter analysis, and
thorough scaling behavior evaluation. Due to the plurality of approaches and stencil patterns, we
set out to develop a generalizable methodology for reproducible measurements accompanied by
state-of-the-art performance models. Our open-source toolchain, and collected results are publicly
available in the ``Intranode Stencil Performance Evaluation Collection'' (INSPECT).

We present the underlying methodologies, models and tools involved in gathering and documenting the
performance behavior of a collection of typical stencil patterns across multiple architectures and
hardware configuration options. Our aim is to endow performance-aware application developers with
reproducible baseline performance data and validated models to initiate a well-defined process of
performance assessment and optimization.

\end{abstract}

\section{Introduction}\label{sec:introduction}

Stencils are relative data access and computation patterns that emerge from the discretization of
differential operators on regular grids. Stencils appear in many fields, from image processing, fluid
dynamics, material sciences to mechanical engineering, and are typically embedded in loop nests that
are at least as deep as the number of dimensions in the original continuous problem. Despite their
apparent simplicity, stencil algorithms show a rich set of performance patterns and allow for
various optimizations in terms of data access and work reduction. For instance, the performance of
most simple stencil algorithms (such as the 3D 7-point constant-coefficient variant encountered with
a simple finite-difference discretization of the Laplace operator on a regular Cartesian grid) is
limited by the memory bandwidth for in-memory working sets on multicore CPUs. Spatial blocking can
reduce the code balance to a theoretical minimum but will not decouple from memory
bandwidth. Temporal blocking can finally render the implementation cache or core bound with
significant performance gains, but there are many different approaches and the number of
parameters is significant~\cite{Malas2017}. Moreover, even recent publications often
fail to assess performance baselines correctly, rendering all reported speedups
meaningless.

\subsection{A Stencil Baseline Performance Collection}

We set out to compile an extensible collection of stencil runtime performance characteristics from a
variety of architectures and backed up by analytic performance models and performance counter
analysis. The stencil update iteration is embedded in a Jacobi algorithm without advanced
optimizations. All resulting information is available in a public online collection, organized based
on an abstract classification scheme, and can be easily reproduced with the presented information
and available open-source tool chain. The collection is also enriched with specific comments on the
performance behavior and model predictions. It can be browsed at
\url{https://rrze-hpc.github.io/INSPECT}.

\subsection{Stencil Classification}\label{sec:classification}

\begin{figure}[b]
    \center
    \includegraphics[width=0.5\columnwidth]{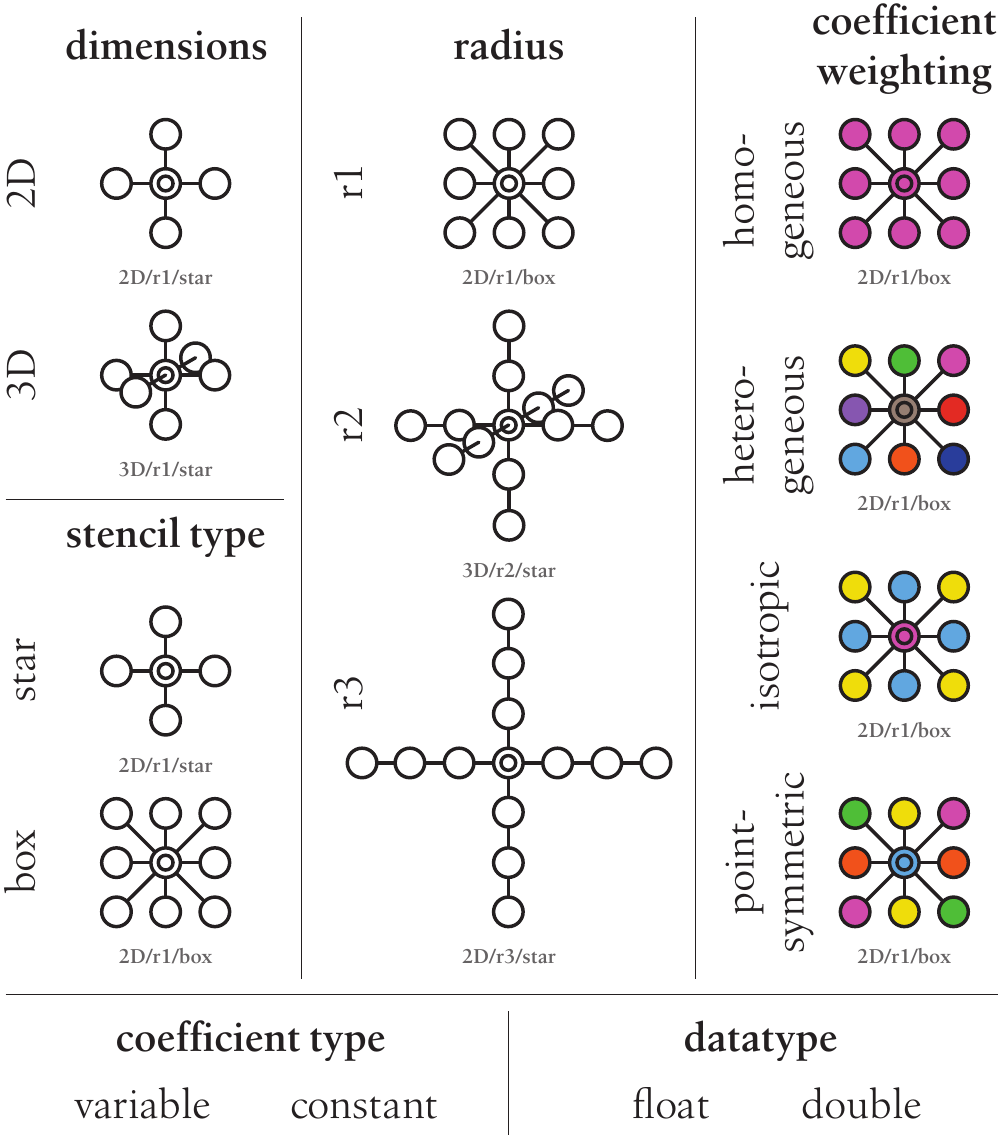}
    \caption{Classification of stencils used by INSPECT. These properties do not capture all aspects, but are representative for a large number of discretizations.}
    \label{fig:stencil_classification}
\end{figure}

In order to span a space of possible stencils we use a classification based on the following
properties:
\begin{itemize}
    \item \emph{dimensions}: dimensionality of the stencil, typically 3D or 2D
    \item \emph{radius}: longest relative offset from the center point in any dimension, usually $r=1$, $r=2$, or $r=3$
    \item \emph{coefficient weighting}: how are coefficients applied, e.g. homogeneous, heterogeneous, isotropic, or point-symmetric
    \item \emph{stencil type}: general shape of stencil, e.g., star or box
    \item \emph{coefficient type}: constant throughout grid or variable at each grid point
    \item \emph{data type}: numeric type of grid elements, e.g., double or float
\end{itemize}
These properties may have a large performance impact depending on the details of the underlying CPU
architecture and its features and configuration, making runtime predictions difficult. Much of the
complexity lies in the cache hierarchy, where data transfers may be handled before they reach main memory and behavior depends on spatial and temporal locality.
On the other hand, in-core bottlenecks such as pipeline latencies, throughput limits,
may also play a decisive role, especially with more complicated stencils.
A visual overview of the stencil classification is given in
Figure~\ref{fig:stencil_classification}. Isotropic coefficient weighting deserves special attention:
All nodes with the same distance to the origin share the same coefficient; different distances have
distinct coefficients.

With the given set of classification properties this leads to at least 192 relevant
combinations. We have not yet gathered data for all possible combinations and architectures
available to us, but a representative set is already available.

This paper is organized as follows: In Section~\ref{sec:background_methodology} we briefly describe
the computer architectural features of the benchmark systems, the analysis tools, and the
performance models we employ to set up the stencil performance
collection. Section~\ref{sec:data_collection} details on our automated workflow and includes a
description of the structure and origin of the data presented on the INSPECT
website. Section~\ref{sec:examples} uses a few distinct stencil examples to showcase the data
presentation and possible insights gained. Finally, Section~\ref{sec:related} gives related work and
Section~\ref{sec:conc} concludes the paper.

\subsection{Contributions}

This work makes the following contributions:
\begin{itemize}
\item a simple classification scheme of stencil characteristics, based on the underlying numerical
  problem, and defining the architecture dependent performance behavior (see
  Sec.~{\ref{sec:classification}}),
\item support for automatic analytic performance model generation for the AMD Zen
  microarchitecture (see Sec.{\ref{sec:examples:zen}}),
\item a first-of-its-kind collection, presentation and method to match the measured
  performance data with automatically generated single- and
  multicore performance models in order to gain insight into relevant performance bottlenecks and
  uncover compiler deficiencies and to guide performance optimization strategies (see
  Sec.~{\ref{sec:data_collection}}, {\ref{sec:examples}} and~{\ref{sec:howto}}),
\item an automatic extraction of the Phenomenological ECM model from hardware
  performance counters (see Sec.~{\ref{sec:bench}}), based on ideas from~\cite{Malas2017},
\item a public website on which the gathered data, performance models, and reproducibility
  information are presented in a clear and structured way (see
  \url{https://rrze-hpc.github.io/INSPECT/}),
\item built-in reproducibility by transparently making all necessary runtime information and
  system configurations available to developers---including the exact commands to execute for reproduction
  (see Sec.~{\ref{sec:examples}}).
\end{itemize}

\section{Background \& Methodology} \label{sec:background_methodology}

To support our methodology, we use STEMPEL~\cite{stempel_github} for code generation
based on the classification shown above, and Kerncraft~\cite{kerncraft} for the generation of
Roofline and Execution-Cache-Memory (ECM) models based on micro-architectural and memory hierarchy
features as well as benchmarking of single- and multi-core scenarios.

We will briefly introduce STEMPEL, Kerncraft and their underlying models ECM and Roofline as well
as hardware features which have a significant impact on similar codes.

\subsection{STEMPEL}

STEMPEL is used to generate stencil codes from the parameters mentioned in
Section~\ref{sec:introduction} (dimensions, radius, coefficient weighting, stencil type,
coefficient type and data type). The resulting kernel code is used as input for Kerncraft, but
STEMPEL also supports generation of benchmarkable code which can be compiled and executed
stand-alone. The generated code may also include OpenMP multithreading support or spatial blocking.
The latter is used to investigate blocking behavior for INSPECT. For accurate extraction of
performance data, the code is additionally instrumented with LIKWID markers to be used with the
\texttt{likwid-perfctr} tool.

\subsection{ECM \& Roofline Model}\label{sec:models}

The Execution-Cache-Memory (ECM)~\cite{Stengel_2015} and Roofline~\cite{Williams_2009} models are
resource-centered performance models that assume certain hardware limitations (such as data
transfer bandwidths, instruction throughput limits, instruction latencies, etc.) and try to map the
application to a simplified version of the hardware in order to expose the relevant bottleneck(s).
This analysis depends on the dataset size and dimensions, as well as the computational effort
during each iteration. Both models generally neglect data access latencies although they can be
added as ``penalties''. Latency predictions would require other models and are usually not of
relevance for stencil code performance. In some cases, latency penalties need to be considered for
``perfect'' predictions~\cite{hofmann17}, but this correction is usually small and is neglected in
this work.

The compute performance bottleneck is analyzed based on the loop body's maximum in-core
performance. Assuming that all load operations will hit the L1 cache, one can estimate the
optimistic runtime the loop body in cycles. The necessary information has been published by Intel,
Agner Fog~\cite{agner_fog}, \url{uops.info}~\cite{Abel19} or through the Intel Architecture Code
Analyzer (IACA)~\cite{IACA}. Although none of those sources are complete, they are good enough for
a well-informed estimation. The resulting inverse throughput of cycles per cacheline (lower is
faster) exposes the bottleneck, for both ECM and Roofline. Cachelines are considered as the
basic work unit, since it is also the basic unit of caches. E.g., for a double precision code with
8\,Byte per element, on a machine with 64\,Byte cachlines, there are eight iterations per
cacheline. For Roofline most publications use performance (higher is faster) as the baseline
metric, but both units can be converted into one another trivially:
\[\mathrm{clock} / \mathrm{performance} \times \mathrm{work} = \mathrm{inverse\ throughput}\]
\[\left[\frac{\mathrm{cycle}}{\mathrm{second}}\right] / \left[\frac{\mathrm{FLOP}}{\mathrm{second}}\right] \times \left[\frac{\mathrm{FLOP}}{\mathrm{iteration}}\right] \times \left[\frac{\mathrm{iteration}}{\mathrm{cacheline}}\right] = \left[\frac{\mathrm{cycle}}{\mathrm{cacheline}}\right]\]
Memory and cache bottlenecks require a prediction of which data access will be served by which
memory hierarchy level. This is either done using a cache simulator (e.g.
\texttt{pycachesim}~\cite{pycachesim}) or the analytical layer-condition model~\cite{kerncraft,
layercondition}. The result from this prediction are the expected traffic volumes between the
levels of the memory hierarchy. The Roofline model then combines the data volume per cacheline
(e.g., eight iterations with double precision) for each hierarchy level with previously measured
bandwidths for the same level and core count, and selects the slowest as the bottleneck. The ECM
model combines all inter-cache transfers with theoretical bandwidths from documentation, and
volumes between memory and last level cache with a measured full-socket bandwidth. These inverse
throughputs are either combined using summation if no overlapping is assumed, maximization for full
overlapping, or any more complicated function for intermediate situations.

For Intel processors, assuming that no overlap between any load, store, inter-cache and memory
transfers happens has proven to be the best fitting model assumption. This might change in future
microarchitectures and does not hold for other vendors.

For the AMD Zen microarchitecture, in-core computations and all inter-cache and register transfers overlap down to the L2 cache. Transfers between L2, L3 and main memory serialize~\cite{hofmann19}.

Another approach is to measure transfers using hardware provided performance counters and base the
ECM model on these empirical volumes and predict the runtime using the non-overlapping assumption.
This is referred to as the \emph{Phenomenological ECM model}, also discussed in Section~\ref{sec:bench}.

The model parameters used by the models are shown in Figure~\ref{fig:architecture} and
Table~{\ref{tab:hosts}}. The corresponding machine files can be found on the INSPECT webpage. For the ECM model
parameters, except for the measured memory bandwidth highlighted by the preceding tilde
(\textasciitilde), all throughputs are published in vendor documentation. The throughputs of
execution ports, scheduler and decoder are not shown here, but most may be found in official
documentation, public resources, or can be benchmarked~\cite{asmbench}. The overall
instruction-level parallelism capability is represented with the different ports.

While in this work the Roofline model is always presented with a single (reciprocal) throughput (TP),
the ECM model is produced from architecture-dependent combinations of in-core computation
TP $T_\mathrm{comp}$, load/store TP $T_\mathrm{RegL1}$, inter-L1/L2 transfer TP
$T_\mathrm{L1L2}$, inter-L2/L3 transfer TP $T_\mathrm{L2L3}$ and main memory transfer TP
$T_\mathrm{L3MEM})$, with the unit of cycles per cacheline.

\subsection{Intel Microarchitectures} \label{sec:architectures}


\begin{table*}
    \caption{Host configuration used for INSPECT. See \url{https://rrze-hpc.github.io/INSPECT/machinefiles} for all details.}
    \label{tab:hosts}
    \footnotesize
    \begin{tabular}{r|cccc}
        \toprule
        Host                      & HSW & BDW & SKX & ZEN \\
        \midrule
        \multirow{2}{*}{CPU model} & Intel & Intel & Intel & AMD \\
         & Xeon E5-2695v3 & Xeon E5-2697v4 & Xeon Gold 6148 & EPYC 7451 \\
        base clock (fixed)        & 2.3 GHz & 2.3 GHz & 2.4 GHz & 2.3 GHz\\
        uncore clock (fixed)      & 2.3 GHz & 2.3 GHz & 2.4 GHz & n/a\\
        turbo mode                & disabled & disabled & disabled & disabled\\
        cores per socket          & 14 & 18 & 20 & 24\\
        cores per NUMA domain     & 7 & 9 & 10 & 6\\
        cluster-on-die (CoD) / sub- & \multirow{2}{*}{CoD enabled} & \multirow{2}{*}{CoD enabled} & \multirow{2}{*}{SNC enabled} & \multirow{2}{*}{n/a}\\
        NUMA-clustering (SNC)     &  &  &  &  \\
        \multirow{2}{*}{microarchitecture} & Haswell & Broadwell & Skylake X & Zen\\
         & see Fig.~\ref{fig:architecture:hsw} & see Fig.~\ref{fig:architecture:hsw} & see Fig.~\ref{fig:architecture:skx} & see Fig.~\ref{fig:architecture:zen} \\
        INSPECT machine name & \tiny \href{https://rrze-hpc.github.io/INSPECT/machines/BroadwellEP\_E5-2697\_CoD}{BroadwellEP\_E5-2697\_CoD} & \tiny \href{https://rrze-hpc.github.io/INSPECT/machines/BroadwellEP\_E5-2697\_CoD}{BroadwellEP\_E5-2697\_CoD} & \tiny \href{https://rrze-hpc.github.io/INSPECT/machines/SkylakeSP\_Gold-6148\_SNC}{SkylakeSP\_Gold-6148\_SNC} & \tiny \href{https://rrze-hpc.github.io/INSPECT/machines/Zen\_EPYC-7451}{Zen\_EPYC-7451}\\
        \bottomrule
    \end{tabular}
\end{table*}

\begin{figure*}
    \subcaptionbox{Haswell (HSW) and Broadwell (BDW)\label{fig:architecture:hsw}}{
        \includegraphics[width=0.5\columnwidth]{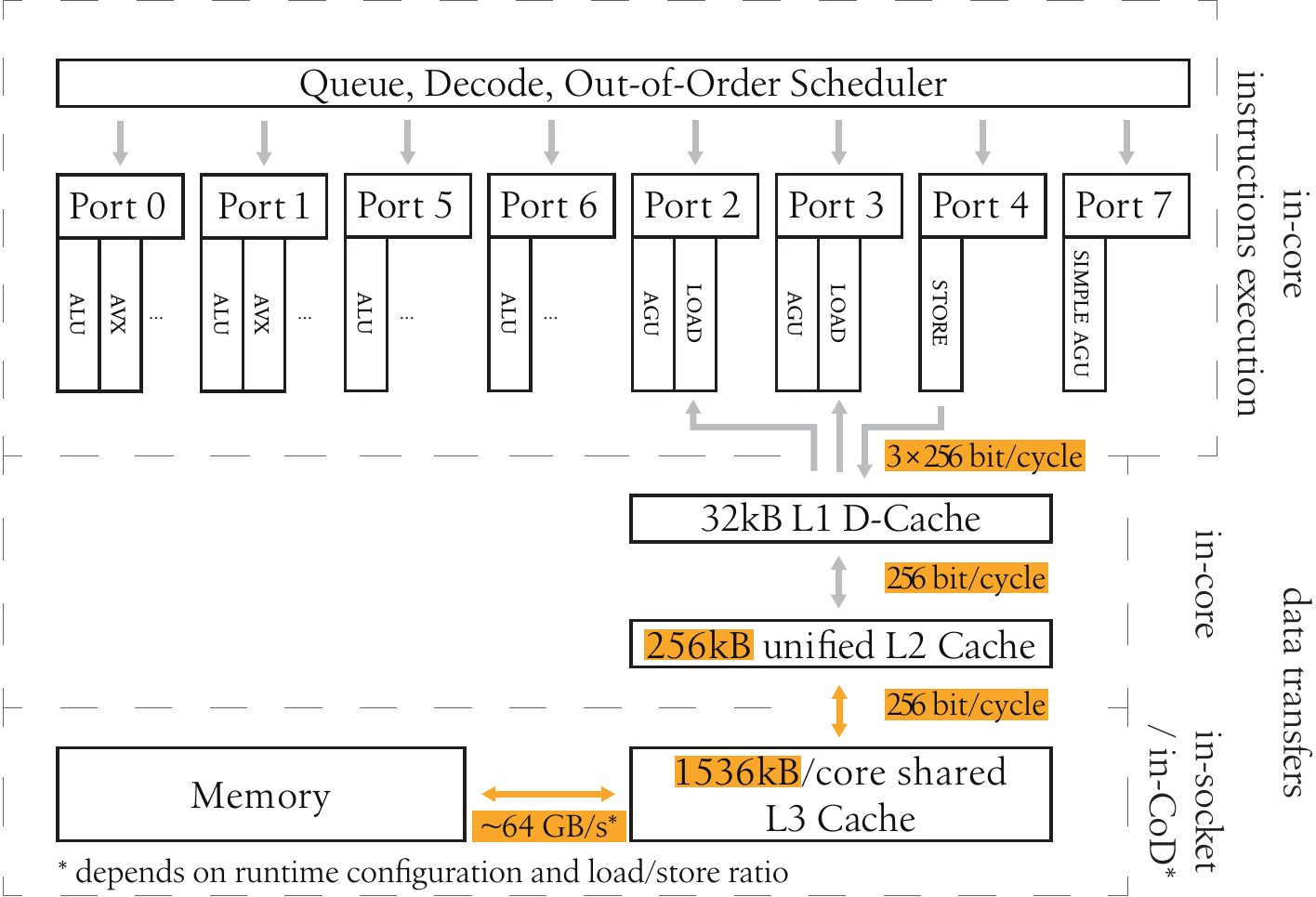}
    }
    \subcaptionbox{Skylake X (SKX)\label{fig:architecture:skx}}{
        \includegraphics[width=0.5\columnwidth]{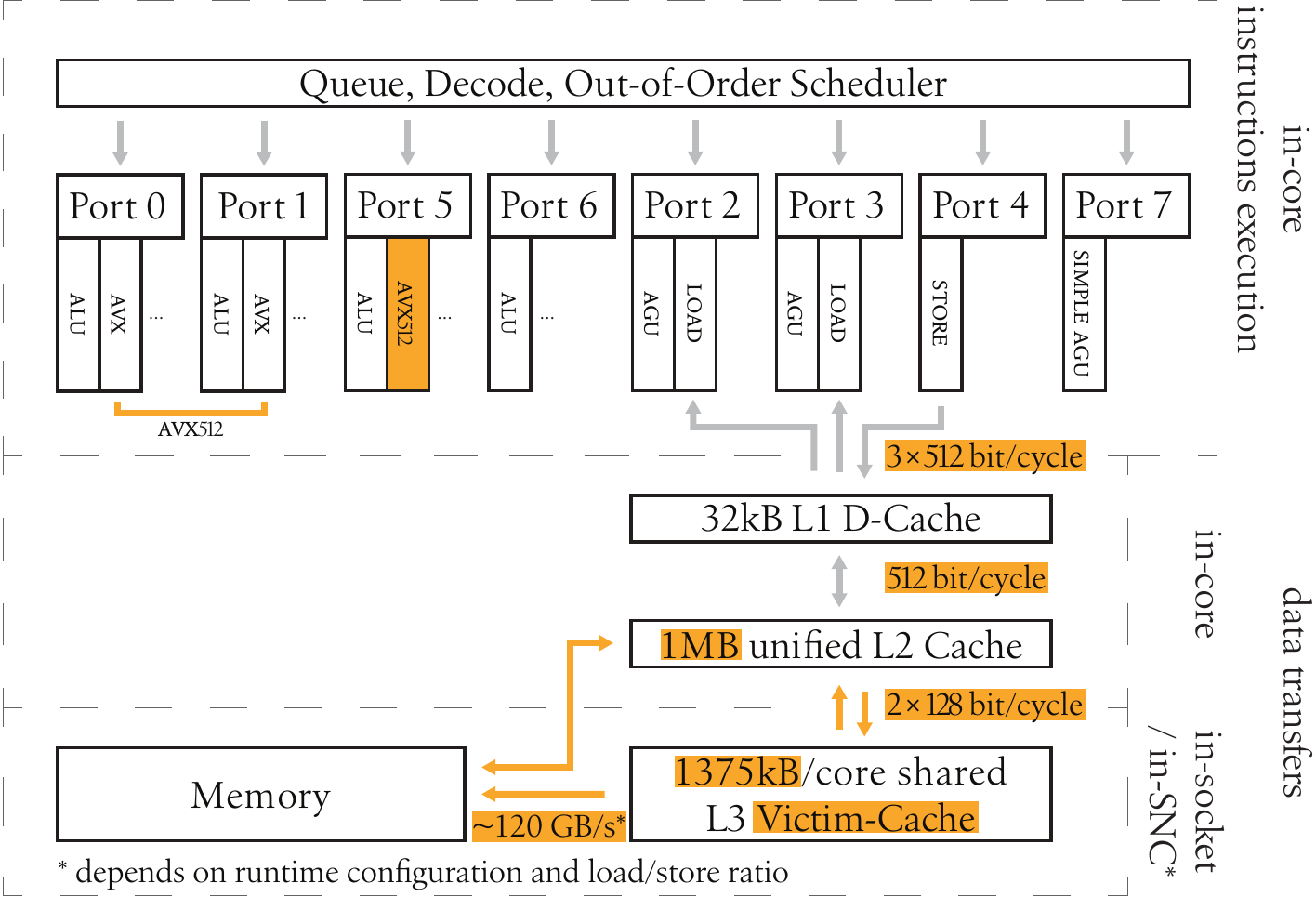}
    }
    \caption{Simplified block diagrams for Intel Haswell, Broadwell and Skylake
      X, including the execution ports and cache
      hierarchy. Differences have been highlighted.
      The Skylake architecture (without X) has no AVX512 support.}
    \label{fig:architecture}
\end{figure*}

The Intel microarchitectures Haswell (HSW) and Broadwell (BDW) have no differences in regard to our
modeling and performance analysis. Figure~\ref{fig:architecture:hsw} shows their architectural
diagram. Both architectures have seven execution ports; most important are the two AVX2
fused-multiply-add (FMA) ports, two load ports able to handle 256-bit per cycle and one 256-bit
store port. AVX and AVX2 instructions can make use of sixteen 256-bit \texttt{YMM} registers. On
the memory side, there is a linear inclusive cache hierarchy.
Due to the double ring interconnect between individual HSW and BDW cores
and separate memory controllers residing on each ring, a cluster-on-die mode can be enabled to
allow NUMA separation of the two rings. With cluster-on-die mode, the last-level
cache (i.e., L3) is split and only used by a core on the same ring and a slightly higher memory bandwidth
can be attained. Results for HSW and BDW are presented with cluster-on-die (CoD) mode
enabled.

The Skylake X (SKX) microarchitecture is shown in Figure~\ref{fig:architecture:skx}. It supports
AVX512, which boosts load, store and FMA ports from 256-bit to 512-bit width. To allow two AVX512
FMA instructions to be executed in parallel (i.e., combined throughput of 0.5 cycles) an AVX512
pipeline was added to Port 5 and the existing 256-bit pipelines at Ports 0 and 1 may be used in
lockstep to reach 2 $\times$ 512-bit width. There are 32 512-bit \texttt{YMM} registers and the
number of 256-bit registers was also doubled to 32. The SKX microarchitecture has a non-inclusive
last level victim cache, which may cache evicted cachelines from L2 and is used to write back to
but not to load from memory. All data coming from memory is loaded directly into the L2 cache of
the requesting core. Unmodified cachelines may also be dropped from L2. The
criteria which decide if a cacheline is evicted from L2 to the victim cache or not have not
been disclosed. For our models we assume that all evicts will be passed on to the last-level cache
and---if changed---stored from there into memory. Sub-NUMA-clustering (SNC) is similar to CoD,
which was also enabled during our measurements.

The changed cache structure of the Skylake microarchitecture with its undocumented decision
heuristic for L3 cache usage and unavailable hardware counters for some of the inter-cache and
memory data paths, still poses a problem. As we will see later, assuming the traditional linear
inclusive cache hierarchy often yields reasonable results, but is still under investigation.

In the architecture diagrams, two-way arrows represent half-duplex capabilities and two individual
arrows means full-duplex capabilities. The factor along with the bandwidth is meant to emphasize
the half- and full-duplex behavior (e.g., between L2 and L3 on Skylake X 128\,bits per cycle may be
transferred both ways concurrently).
This information and more is used to construct a machine file for Kerncraft and will be explained
in Section~\ref{sec:machine_model}.

The specific systems that are used for INSPECT are documented at \url{https://rrze-hpc.github.io/INSPECT/machinefiles}. A summary of the relevant configuration details, in addition to the micro architectural details in Figure~\ref{fig:architecture}, is given in Table~\ref{tab:hosts}.

\subsection{AMD Zen Microarchitecture}
\begin{figure}
    \center
    \includegraphics[width=0.7\columnwidth]{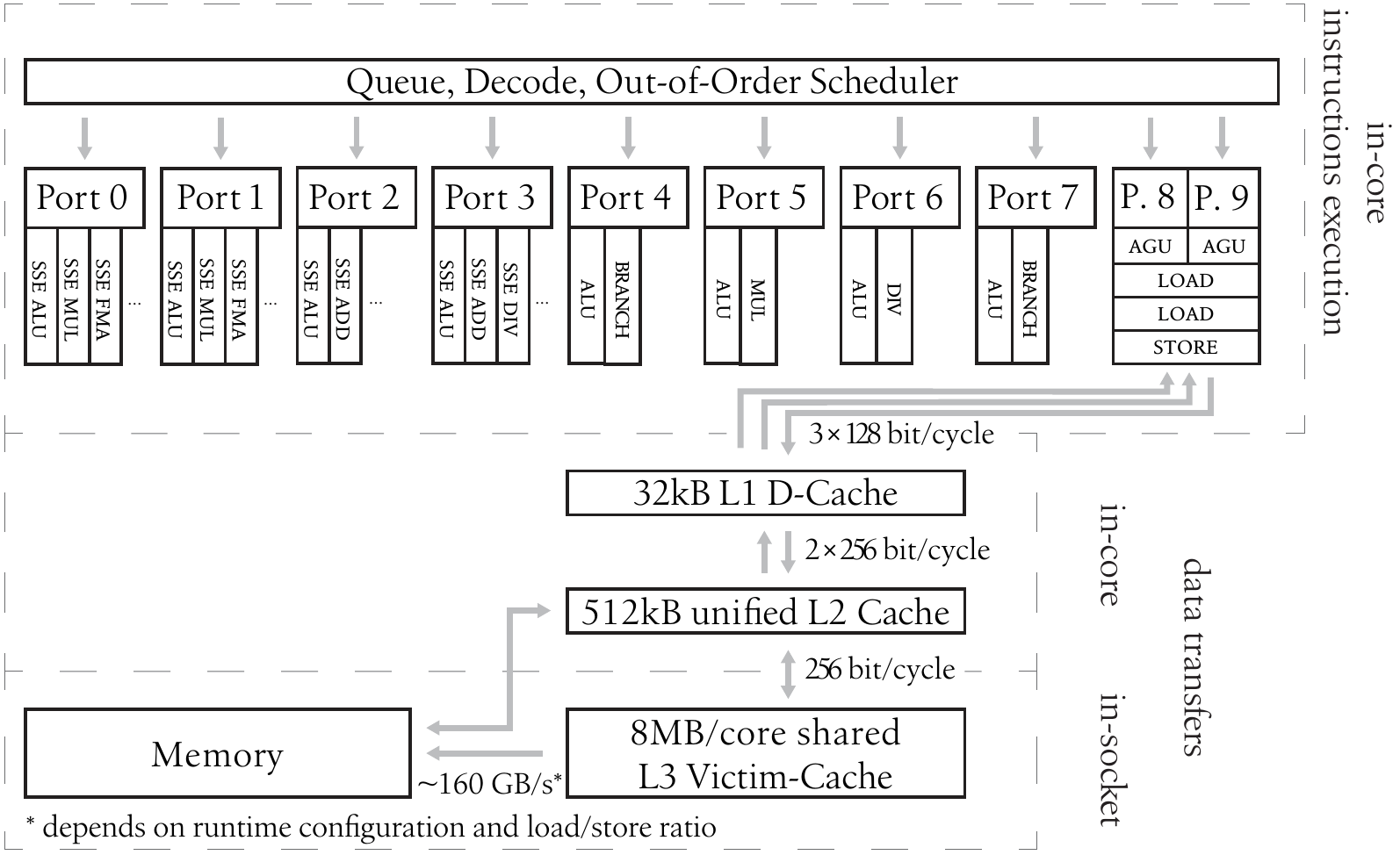}
    \caption{\label{fig:architecture:zen}Simplified block diagram for AMD Zen microarchitecture, including execution ports and cache hierarchy.}
\end{figure}

The AMD Zen microarchitecture has ten ports, the first four of which (0, 1, 2 and 3)
support 128-bit wide floating-point SSE instructions (see Figure~{\ref{fig:architecture:zen}}).
Each execution unit, except for divide, is present on two
ports, e.g., FMA and MUL on 0 and 1, and ADD on 2 and 3. The decoder stage supports AVX instructions
by utilizing two SSE ports simultaneously (similar to AVX512 on Port 0 an 1 with Skylake X). Ports
4 through 7 handle integer and control flow instructions.

Ports 8 and 9 each have their own address generation unit (AGU) and can utilize the two shared
load ports and the single shared store port. The store and load ports each operate on up to 128 bits
and can issue one load/store to the 32\,kB first-level (L1) cache. Thus either two
loads or one load and one store can be executed per cycle at maximum. The 512\,kB inclusive L2
cache is connected to the L1 cache with 256-bit per cycle full-duplex (i.e., a 64\,B cacheline
takes two cycles to transfer, and loads and stores are done in parallel). Between L2 and the
last-level cache (L3), 256 bits can be transferred per cycle, but only half-duplex (i.e.,
either load or store). The exact heuristics of the victim cache are not publicly available. In
addition to that, support for hardware performance counters is much more limited, which does not
allow us to inspect many of the transfers between memory levels.

The maximum main memory bandwidth achievable is close to 160\,GB/s, 30\% higher than what we were
able to measure on Skylake X. The specific AMD CPU used here has 24 cores, split over four NUMA domains of 6 cores each.

\subsection{Kerncraft}


Kerncraft brings together static analysis of the kernel code with microarchitecture data and
execution models into a coherent performance prediction model, based on the Roofline and ECM
models. It also allows benchmarking of the kernel codes with single and multiple cores (using
OpenMP) and collection of hardware performance counter data during execution.

\subsubsection{Machine Model} \label{sec:machine_model}

The specific machine model for each microarchitecture is described in the \emph{machine files},
which are provided either with Kerncraft or can be generated semi-automatically using
\texttt{likwid\_bench\_auto}---a tool distributed with Kerncraft. All machine models mentioned in
this paper are provided with Kerncraft in the examples directory. When using the semi-automatic
generation, it must be executed on such a machine and the resulting file needs to be completed
manually from vendor documentation, model assumptions or from existing---similar---architecture
files.

Machine model files contain detailed information on the architecture and memory hierarchy as well
as benchmark results, necessary to construct the Roofline and ECM model. In particular
STREAM~\cite{McCalpin1995} benchmark results, cache sizes and parameters, NUMA topology, base clock
and architecture specific compiler arguments make up the majority of the description. Some of it
can be collected automatically, some needs to be provided manually.

The INSPECT website presents a breakdown of the architecture information in the machine files:
\url{https://rrze-hpc.github.io/INSPECT/machinefiles}. There may be known issues, which are also
documented on INSPECT. E.g., Haswell's L1-L2 bandwidth is theoretically 64\,Byte per cycle, but
benchmarks show that the achievable bandwidth may be as low as 32\,B/cy. Kerncraft assumes the
optimistic 64\,B/cy.

\subsubsection{Model Construction}

The static analysis and model building is split in two parts: in-core execution and data analysis;
both of which are done without execution of the kernel code and can be performed on any hardware.

The in-core analysis is done via the Intel Architecture Code Analyzer tool (IACA)~\cite{IACA} for
Intel architectures and the Open Source Architecture Code Analyzer (OSACA)~\cite{OSACA},
which yields the number of cycles each execution port is occupied by the kernel's assembly
instructions ($T_\mathrm{comp}$ and $T_\mathrm{RegL1}$ for the ECM model). Kerncraft takes care of
compilation, unrolling and vectorization in order to correctly interpret the IACA/OSACA result and relate it to high-level loop iterations found in the kernel source code.

The data analysis predicts inter-cache and memory transfer data volumes either using the analytical
layer-condition model (LC) or the pycachesim~\cite{pycachesim} cache simulator. This yields
$T_\mathrm{L1L2}$, $T_\mathrm{L2L3}$ and $T_\mathrm{L3MEM}$ for the ECM model. The LC model
analysis is very fast and gives a closed form analytical model, but relies on an idealistic
full-associative inclusive and least-recently-used cache hierarchy. The cache simulator can handle
more realistic and complex cache configurations, such as associativity and non-inclusive cache
hierarchies---at the cost of speed and without a closed form solution. Certain aspects of real
hardware can not be simulated due to missing documentation, e.g., cache placement algorithm for
last-level cache on current Intel microarchitectures.

For multi-core scaling, we use the memory latency penalty estimation as described by
Hofmann~\cite{Hofmann_isc18}.

\subsubsection{Benchmark Mode and Phenomenological ECM} \label{sec:bench}

Benchmarking of any code can be tricky, this is the same for stencil codes. Kerncraft takes care of
pinning and hardware performance counter monitoring with LIKWID~\cite{likwid}, as well as ensuring
a minimal runtime, checking machine configuration and derivation of relevant metrics from the
measurements. The underlying performance counters are defined in the machine model and based on
validated metrics provided by LIKWID.

In addition, metrics based on measurement of the runtime, such as memory bandwidths and lattice
updated per second, data transfer volumes can be measured accurately. From these Kerncraft can
construct a \emph{Phenomenological ECM model}. This phenomenological model is not based on the
measured runtime or derived bandwidths, but uses inter-cache and memory data volumes as well as
counts of executed $\mu$ops per port. The overall prediction is then compiled in the same way as
the analytical ECM prediction is compiled from vendor documented transfer rates, measured memory
bandwidth and instruction throughput information.

The specific counters necessary have been compiled from Intel documentation and their
correctness validated with microbenchmarks, where possible. This process is part of the ongoing
LIKWID development. Kerncraft's machine models put the counter in relation to ECM model parameters,
such as L1-L2 traffic or execution port utilization. To measure all necessary counters, multiple
executions are unavoidable, because only a limited number of counter registers are available for
use. On Intel's server microarchitectures, many performance counters are available and a complete
model can be assembled as presented in Figure~{\ref{fig:3Dr1homostarconstdoubleHSW}f}.
On AMD Zen, however, essential contributions, such as main memory traffic can not be examined and a
complete phenomenological model may not be constructed.
\section{Data Collection} \label{sec:data_collection}
\begin{algorithm}
\caption{INSPECT data collection workflow\label{alg:INSPECT-datacollection}}
\begin{algorithmic}[0]
\State fix frequency
\For{dimension, radius, kind, coefficients, weighting, data type}
    \State \texttt{stempel}: generate stencil
    \State \texttt{kerncraft}: layer condition analysis
    \State $\texttt{LC}_{\texttt{L3,3D}} \gets $ compute L3 3D layer condition size
    \State $\texttt{N}_{\texttt{L3,3D'}} \gets 1.5 \cdot \texttt{LC}_{\texttt{L3,3D}}$
    \If{$\texttt{N}_{\texttt{L3,3D'}} < $ available Memory per NUMA-domain}
        \State reduce \texttt{grid\_size} until it fits in memory
    \EndIf
    \For{$n \gets 10, \texttt{N}_{\texttt{L3,3D'}}$}
        \Comment{single core grid scaling}
        \State \texttt{kerncraft}: (LC, CS)$\,\times\,$(RooflineIACA, ECM)
        \State \texttt{kerncraft}: Benchmark
    \EndFor
    \For{$Threads \gets 1, \texttt{N}_{threads}$}
        \Comment{thread scaling @ $\texttt{N}_{\texttt{L3,3D'}}$}
        \State \texttt{kerncraft}: (RooflineIACA, ECM, Benchmark)
    \EndFor
    \State \texttt{stempel}: generate 3D spatial blocking code
    \For{$n \gets 10, \texttt{N}_{\texttt{L3,3D'}}$}
        \Comment{single core grid scaling}
      \State determine 'good' 3D blocking factors
        \State \texttt{likwid-perfctr}: Benchmark
    \EndFor
    \State csv data + graphs + website \Comment{post processing}
\EndFor
\end{algorithmic}
\end{algorithm}

In order to build a comprehensive single-node stencil performance database, preexisting open-source
tools STEMPEL~\cite{stempel_github}, Kerncraft~\cite{kerncraft} and LIKWID~\cite{likwid} have been
combined in the ``Intranode Stencil Performance Evaluation Collection'' (INSPECT). For given
stencil parameters all benchmark and automated performance modeling data for the present machine
can be collected with a single (job) script. The data collection work flow can be seen in
Algorithm~\ref{alg:INSPECT-datacollection}.

For the stencil source code generation, to be supplied to kerncraft, STEMPEL is used. Possible
parameters are: dimension, radius, stencil type, coefficient weighting and type as well as the
data type. Examples of the produced stencil code by STEMPEL are shown in
Listings~\ref{lst:3D-1r-homo-star-const-double},~\ref{lst:3D-3r-hetero-star-const-double}
and~\ref{lst:3D-1r-hetero-box-const-double}. If a custom stencil is to be be used, this step can be
omitted.

The stencil code is then supplied to Kerncraft in order to do layer condition analysis and
determine sensible ranges for grid sizes to be examined. Data ranges are chosen such that the last
level cache 3D layer condition is violated and a steady state will be reached as long as the
available main memory per NUMA domain is not exceeded (see $1.5 \cdot LC_{L3,3D}$
in Algorithm~\ref{alg:INSPECT-datacollection}).

The next step is data collection. For single core grid scaling, Kerncraft is used to generate
Roofline and ECM performance models with layer conditions and cache simulation, as well as
benchmark and Phenomenological ECM data. Multi-core thread scaling is done for the largest
previously calculated, memory bound grid size for all cores of one socket. Here Kerncraft is again
used to generate Roofline, ECM and Benchmark data. In a last step spatial blocking is performed.
Here STEMPEL is once more used to generate executable benchmark code with spatial blocking from the
basic stencil code, generated before. This spatial blocking code is then instrumented with LIKWID
to obtain the required benchmark data.

In a final step all data is collected, postprocessed and archived. The outputs are the data files
that are needed for the visualization on the website. Those files can be pushed to the git
repository to automatically include the inspected stencil on the INSPECT website.

For every stencil-machine configuration the website shows general stencil information, graphs of
the measured and predicted performance and step-by-step instructions for the replication of the
shown data.
The general stencil information contains: stencil parameters, kernel source code, kernel assembly
and layer condition analysis, as well as IACA throughput analysis and information about the state
of the machine and operating system, the data was collected on.
Performance prediction and benchmark data are shown in 5 different graph types:
\begin{itemize}
  \item stacked ECM (with layer condition, cache simulation and phenomenological)
  \item Roofline performance (with layer condition and cache simulation)
  \item data transfers for single core grid scaling
  \item full-socket thread scaling (one grid size by default, but possibly more)
  \item spatial blocking performance plots (3D-L3 cache blocking by default, but possibly more)
\end{itemize}
An example of the plots visible for each stencil configuration is shown in
Figure~\ref{fig:3Dr1homostarconstdoubleHSW}.
The reproducibility information contains detailed steps on how to generate the stencil code with
STEMPEL and all necessary commands to retrieve the data shown on the site.

Additionally all shown data can be commented and validated with a traffic light system reflecting
the quality of the shown plots. This allows to highlight problems or unintuitive results of a
specific stencil or hardware configuration, that could otherwise be mistaken for incorrect data.

\section{Examples}\label{sec:examples}

Three different exemplary stencil configurations were selected from the INSPECT website:
short-/long-ranged and star/boxed stencils, on three different machines. We will start with a very
basic 7-point stencil on a Haswell Xeon E5-2695v3 machine, then continue with a long-ranged stencil
on Skylake Xeon Gold 6148 and compare a boxed stencil on Broadwell Xeon E5-2697v4 and Skylake Xeon
Gold 6148; we will conclude with the basic 7-point stencil on an AMD EPYC 7451 machine.

In addition to the presented architectures, the INSPECT website also contains analyses and
measurements on Intel Sandy Bridge and Ivy Bridge architectures.

\subsection{A Simple Short-ranged Stencil on Haswell (Intel Xeon E5-2695v3)}\label{sec:example:shortrange}

\begin{figure*}
  \centering
  \input{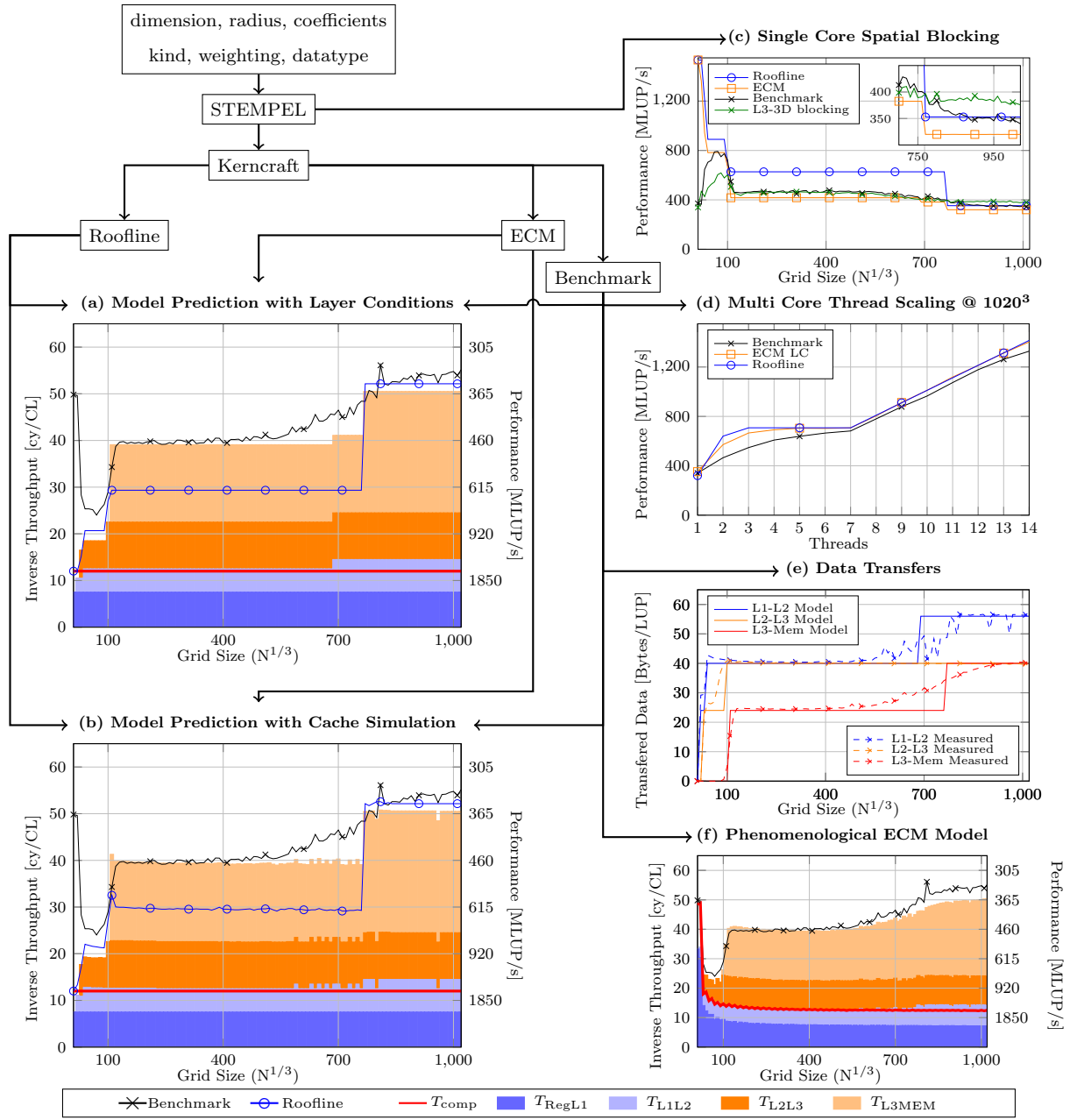}
  \caption{Flow chart visualizing the connections between STEMPEL and Kerncraft and the resulting
  data. STEMPEL is used to generate the kernel and spatial blocking code. Kerncraft performs
  Roofline and ECM analysis on the kernel code for layer condition and cache simulation.
  Kerncraft's benchmark mode provides the measurement data for grid scaling, data transfers and
  Phenomenological ECM, as well as multi core thread scaling. The shown data is for a 3D star
  stencil with radius 1 and constant, homogeneous coefficients with double precision floating-point
  numbers on a Haswell Xeon E5-2695v3 machine with Cluster on Die mode enabled. Underlying data,
  plots and additional model information may be viewed at \url{https://git.io/fjqzy} on the INSPECT
  website.}
  \label{fig:3Dr1homostarconstdoubleHSW}
\end{figure*}


\begin{lstlisting}[language=C,float,floatplacement=H,label={lst:3D-1r-homo-star-const-double},caption={3D stencil code with radius 1, constant homo-geneous coefficients, star structure and double data type.}]
double a[M][N][P], b[M][N][P], c0;

for(long k = 1; k < M - 1; ++k) {
  for(long j = 1; j < N - 1; ++j) {
    for(long i = 1; i < P - 1; ++i) {
      b[k][j][i] = c0 * ( a[k][j][i]
                        + a[k][j][i-1] + a[k][j][i+1]
                        + a[k][j-1][i] + a[k][j+1][i]
                        + a[k-1][j][i] + a[k+1][j][i] );
}}}
\end{lstlisting}

The first stencil configuration presented here is the simple 3D 7-point stencil on the well
understood Haswell microarchitecture (Intel Xeon E5-2695v3 in CoD mode): 3D, radius 1, star
stencil, constant and homogeneous coefficients with double precision floating-point accuracy. The
stencil source code is shown in Listing~\ref{lst:3D-1r-homo-star-const-double}.
Figure~\ref{fig:3Dr1homostarconstdoubleHSW} displays all graphs presented on the INSPECT website,
as well as the workflow from stencil generation with STEMPEL to data acquisition with Kerncraft, as
already outlined by Algorithm~\ref{alg:INSPECT-datacollection}. The model prediction graphs show a
stacked ECM prediction ($T_\mathrm{ECM} = \max(T_\mathrm{comp}, \mathrm{sum}(
T_\mathrm{RegL1}, T_\mathrm{L1L2}, T_\mathrm{L2L3},
T_\mathrm{L3MEM}))$), together with the Roofline prediction and benchmark measurement data.

All the presented data and plots on this specific kernel, including commands to reproduce, system
configuration used and very verbose information on the analysis (such as the IACA output) may be
viewed at: \url{https://git.io/fjqzy}.

Figure~\ref{fig:3Dr1homostarconstdoubleHSW}a shows the ECM and Roofline prediction generated by
Kerncraft, based on the layer condition prediction. Figure~\ref{fig:3Dr1homostarconstdoubleHSW}b is
based on the cache simulation. The stacked colored regions represent the data transfers in the ECM
model, where the upper boundary is equal to the ECM prediction including all data transfers
($T_\mathrm{RegL1} + T_\mathrm{L1L2} + T_\mathrm{L2L3} +
T_\mathrm{L3MEM}$). The red line represents the compute bottleneck
($T_\mathrm{comp}$), as predicted with IACA. The Roofline prediction is the thin blue line
with circles. The black line with x's are the measurement results from running Kerncraft in
benchmark mode.

The Roofline prediction is accurate when all layer conditions are violated and all stencil data is
coming from main memory. Before that, its prediction is about 25\% too optimistic. In the
  transition zone, there are a few points where the Roofline model is too pessimistic, because the mathematical layer condition is sharp but the performance shows a smooth transition because of the cache replacement policy. The ECM model results in a much more
accurate prediction. All layer condition jumps ($30^3$: L1-3D, $90^3$: L2-3D, $680^3$: L1-2D,
$760^3$: L3-3D) are clearly visible and correspond to the measured performance. The large deviation
between models and measurement in the initial section ($N<100^3$) comes from loop overheads and
large impacts of remainder loop iterations, while this is expected, it is not modeled by neither
ECM nor Roofline.

Comparing the cache simulator (Fig.~\ref{fig:3Dr1homostarconstdoubleHSW}b) with the layer
conditions (Fig.~\ref{fig:3Dr1homostarconstdoubleHSW}a) shows that some peaks or dips in the
measurement can be explained by using a more accurate cache model as provided by the cache
simulator. It also allows for a smoother transition between broken layer conditions, but
nonetheless fails at accurately predicting the transition behavior. Perfect tracking of those
transitions, as seen in the benchmark measurements, would only be possible with precise knowledge
of the underlying caching algorithms implemented in the different cache levels. Due to a lack of
information from the CPU vendors a perfect LRU cache is assumed, as well as other idealized
implementation details.

In the Phenomenological ECM graph, cf. Figure~\ref{fig:3Dr1homostarconstdoubleHSW}f, the smooth
transition between broken layer conditions can be tracked very well. Apart from the transitions
zones, the individual contributions as modeled by the analytical ECM model
((Fig.~\ref{fig:3Dr1homostarconstdoubleHSW}a and \ref{fig:3Dr1homostarconstdoubleHSW}b) and derived
from measurements in the Phenomenological ECM model match up very well. It also shows why the
measured performance differs immensely from the predictions below $100^3$, which now shows as very
high in-core execution time an load instructions because of short scalar loops. The difference
between measurement and model towards the right side of the graph hints that there are saturation
effects in the memory interface which we do not yet understand fully.

The data transfer volumes predicted by the layer conditions and their comparison with measured data
volumes through hardware performance counters can be seen in
Figure~\ref{fig:3Dr1homostarconstdoubleHSW}e. The solid lines show the data transfer prediction
between cache levels and main memory and the dashed lines are the measured data transfers. Between
$100^3$ and $500^3$ as well as after $850^3$ the predicted transfers at each level and the measured
data fit perfectly and show the accuracy of this method. As layer condition break the measured
cache transfers show a smooth transition, until it realigns with the predicted data volume.

Figure~\ref{fig:3Dr1homostarconstdoubleHSW}c shows the impact of cache blocking for specific
layer conditions. In this case, blocking was performed for the L3-3D layer condition, where only
the middle loop (e.g., \texttt{j}-loop in Listing~{\ref{lst:3D-1r-homo-star-const-double}}) is
blocked to keep the relevant data at least in the L3 cache and reduce main memory traffic to a
minimum. As intended, performance stays constant after the L3-3D layer condition is broken, with
spatial cache blocking enabled (green line). This behavior can be predicted from
Fig.~{\ref{fig:3Dr1homostarconstdoubleHSW}}a, where spatial blocking means to preserve the
throughput of an earlier plateau while increasing the dataset size. Reasonable blocking factors are
given by the range of the plateau (e.g., here the block dimensions should be
$N_\mathrm{block}^{1/3} < 700$). Blocking for the next lower plateau (i.e., $N_\mathrm{block}^{1/3}
< 100$) may introduce too much overhead due to short loops. Another, more complicated option would
be the use of temporal blocking, which is expected to yield about the same performance as
$N_\mathrm{block}^{1/3} < 100$, because stripping the top contribution from the stacked plot would
bring the throughput to the same plateau.

Moving on from the single core to multi core scaling, Figure~\ref{fig:3Dr1homostarconstdoubleHSW}d
shows the in-socket thread scaling behavior at $1020^3$. Due to the Cluster-on-Die configuration of
the machine, the performance flattens out at the end of the first NUMA-domain (7 cores). With the
addition of the second NUMA-domain a linear increase can be seen, due to the linear addition of
bandwidth from the added cores based on the compact scheduling scheme. Predictions of ECM and
Roofline models fit very well in the second, linear part of the graph, and the ECM is also able to
capture the phase before memory bandwidth saturation.

\subsection{A Simple Short-ranged Stencil on AMD Zen} \label{sec:examples:zen}

In Figure~\ref{fig:3D1rhomostarconstdoubleZEN} we show an analysis on the AMD Zen
microarchitecture, presenting results for the same kernel as in
Listing~\ref{lst:3D-1r-homo-star-const-double}. These and additional results may be found at
\url{https://git.io/fj4yq}. As described in Section~\ref{sec:models}, the AMD Zen
architecture shows strong overlap in data transfers. The port execution model is based
on the OSACA implementation~\cite{OSACA}, and the Kerncraft version used for this is based on the
latest \texttt{feature/OSACA} branch. For data volume prediction we use the layer condition model, as for SKX.

The ECM prediction for the AMD Zen microarchitecture is based on the following model, which has
fewer serializing terms: \[T_\mathrm{ECM} = \max(T_\mathrm{comp}, T_\mathrm{RegL1},
T_\mathrm{L1L2}, \mathrm{sum}(T_\mathrm{L2L3}, T_\mathrm{L3MEM}))\] This difference is also visible
in Figure~{\ref{fig:3D1rhomostarconstdoubleZEN}}, where the overlapping parallel terms
($T_\mathrm{comp}$, $T_\mathrm{RegL1}$ and $T_\mathrm{L1L2}$) are simple lines and the serial
contribution terms ($T_\mathrm{L2L3}$ and $T_\mathrm{L3MEM}$) are stacked onto one another.

As with Skylake X, the benchmark follows the trend of the model qualitatively, but measurements
yield better throughput with increased main memory traffic. This effect is seen in both the ECM and
the Roofline model and we believe it is linked to the undisclosed behavior of the L3 cache. The cache
simulator apparently overestimates the number of L2 or L3 misses and predicts a higher main memory
traffic volume. Unfortunately, AMD Zen does not have main memory traffic hardware performance
counters, so we are unable to validate this assumption.

In light of the large main memory traffic contribution, we would suggest temporal blocking to bring
the inverse throughput down to the $T_\mathrm{RegL1}$ level. $T_\mathrm{L3Mem}$ contains all memory accesses (i.e., transfers between main memory, L3 and L2).

\begin{figure*}
  \begin{minipage}[t]{0.48\textwidth}

\pgfplotstableread[col sep=comma]{Plots/3D-1r-homo-star-const-double-ZEN.csv}\datatwo

\pgfplotsset{ECM_front_style/.style={
      width=8.25cm,
      height=6cm,
      xlabel={Grid Size (N$^{1/3}$)},
      ylabel={Inverse Throughput [cy/CL]},
      title style={yshift=.5cm},
      every axis y label/.style={at={(current axis.west)},left=1.25cm,rotate=90,font=\scriptsize},
      every axis x label/.style={at={(current axis.south)},yshift=-.75cm,font=\scriptsize},
      grid=major,
      ymin=0,ymax=50,
      ytick={0,10,...,40},
      ytick pos=left,
      xmin=10,xmax=760,
      xtick={100,300,...,700},
      mark repeat=10,
      tick label style={font=\footnotesize},
      label style={font=\small},
      ylabel style={anchor=north}
}}

\pgfplotsset{ECM_back_style/.style={
      width=8.25cm,
      height=6cm,
      xlabel={},
      ylabel={},
      every axis y label/.style={at={(current axis.west)},left=1.25cm,rotate=90,font=\scriptsize},
      every axis x label/.style={at={(current axis.south)},yshift=-.75cm,font=\scriptsize},
      ymin=0,ymax=50,
      ytick={0,10,...,40},
      ybar stacked,
      bar width=2.25pt,
      xmin=10,xmax=760,
      xtick={100,300,...,700},
      tick label style={font=\footnotesize,white},
      label style={font=\small},
      ylabel style={anchor=north}
}}

\pgfplotsset{ECM_secondary_axis_style/.style={
      ECM_front_style,
      title={},
      grid=none,
      axis y line*=right,
      axis x line=none,
      ymin=0, ymax=1000,
      ytick={200,400,...,800},
      yticklabels={460,610,920,1840},
      ytick pos=right,
      y dir=reverse,
      ylabel={Performance [MLUP/s]},
}}

\begin{tikzpicture}[scale=.85]
    \begin{axis}[ECM_back_style,at={(0,0)}]
        \addplot +[ybar,orange,mark=none] table [x = {N^3},y = {ECM LC Tl2l3}] {\datatwo};
        \addplot +[ybar,orange!50!white,mark=none] table [x = {N^3},y = {ECM LC Tl3mem}] {\datatwo};
    \end{axis}

    \begin{axis}[ECM_front_style,at={(0,0)}]
        \addplot+ [mark=x,black] table[x = {N^3},y = {Benchmark cycl}]{\datatwo};
        \addplot+ [mark=square,red, mark size=2] table[x = {N^3},y = {ECM LC Tol}]{\datatwo};
        \addplot +[mark=star,blue!60!white, mark size=2] table [x = {N^3},y = {ECM LC Tnol}] {\datatwo};
        \addplot +[mark=triangle,blue!30!white, mark size=2] table [x = {N^3},y = {ECM LC Tl1l2}] {\datatwo};
        \addplot+ [mark=o,blue] table[x = {N^3},y = {Roofline LC cycl}]{\datatwo};
    \end{axis}

    \begin{axis}[ECM_secondary_axis_style,at={(0,0)},
      every axis y label/.style={at={(current axis.east)},right=2cm,rotate=-90,font=\small,xshift=-1.5cm}]
    \end{axis}

    \path [-,thick,draw] (-1.1,4.75) -- (-.85,4.75);
      \node at (-.975 ,4.75) {\tiny$\times$};
      \node [anchor=west] at (-.95 ,4.75) {\tiny Benchmark};
    \path [-,thick,draw,blue] (.75,4.75) -- (1.05 ,4.75);
      \node [blue] at (.9,4.7375) {\large$\circ$};
      \node [anchor=west] at (.9,4.75) {\tiny Roofline};
    \path [-,thick,draw,red] (2.25 ,4.75) -- (2.45 ,4.75);
      \node [red] at (2.35,4.725) {\tiny$\Box$};
      \node [anchor=west] at (2.3,4.75) {\tiny $T_{\scaleto{\mathrm{comp}}{3pt}}$};
    \path [-,thick,draw,blue!60!white] (3.3 ,4.75) -- (3.5 ,4.75);
      \node [blue!60!white] at (3.4,4.7625) {\scriptsize$\star$};
      \node [anchor=west] at (3.35 ,4.75) {\tiny $T_{\scaleto{\mathrm{RegL1}}{3pt}}$};
    \path [-,thick,draw,blue!30!white] (4.3 ,4.75) -- (4.5,4.75);
      \node [blue!30!white] at (4.4,4.75) {\tiny$\triangle$};
      \node [anchor=west] at (4.4,4.75) {\tiny $T_{\scaleto{\mathrm{L1L2}}{3pt}}$};
    \path [-,line width=.2cm,draw,orange] (5.4 ,4.75) -- (5.6,4.75);
      \node [anchor=west] at (5.5,4.75) {\tiny $T_{\scaleto{\mathrm{L2L3}}{3pt}}$};
    \path [-,line width=.2cm,draw,orange!50!white] (6.55 ,4.75) -- (6.75 ,4.75);
      \node [anchor=west] at (6.65 ,4.75) {\tiny $T_{\scaleto{\mathrm{L3Mem}}{3pt}}$};
\end{tikzpicture}
    \caption{\label{fig:3D1rhomostarconstdoubleZEN} ECM and Roofline predictions with benchmark results for 3D 7-point star stencil on AMD ZEN.}
  \end{minipage}~~~
  \begin{minipage}[t]{0.48\textwidth}

\pgfplotstableread[col sep=comma]{Plots/3D-3r-hetero-star-const-double-SKX_SNC.csv}\datatwo

\pgfplotsset{ECM_front_style/.style={
      width=8.25cm,
      height=6cm,
      xlabel={Grid Size (N$^{1/3}$)},
      ylabel={Inverse Throughput [cy/CL]},
      title style={yshift=.5cm},
      every axis y label/.style={at={(current axis.west)},left=1.25cm,rotate=90,font=\scriptsize},
      every axis x label/.style={at={(current axis.south)},yshift=-.75cm,font=\scriptsize},
      grid=major,
      ymin=0,ymax=110,
      ytick={0,20,...,100},
      ytick pos=left,
      xmin=10,xmax=1020,
      xtick={100,400,...,1000},
      mark repeat=10,
      tick label style={font=\footnotesize},
      label style={font=\small},
      ylabel style={anchor=north}
}}

\pgfplotsset{ECM_back_style/.style={
      width=8.25cm,
      height=6cm,
      xlabel={},
      ylabel={},
      every axis y label/.style={at={(current axis.west)},left=1.25cm,rotate=90,font=\scriptsize},
      every axis x label/.style={at={(current axis.south)},yshift=-.75cm,font=\scriptsize},
      ymin=0,ymax=110,
      ytick={0,20,...,100},
      ybar stacked,
      bar width=1.75pt,
      xmin=10,xmax=1020,
      xtick={100,400,...,1000},
      tick label style={font=\footnotesize,white},
      label style={font=\small},
      ylabel style={anchor=north}
}}

\pgfplotsset{ECM_secondary_axis_style/.style={
      ECM_front_style,
      title={},
      grid=none,
      axis y line*=right,
      axis x line=none,
      ymin=0, ymax=1100,
      ytick={100,300,...,900},
      yticklabels={190,240,320,480,960},
      ytick pos=right,
      y dir=reverse,
      ylabel={Performance [MLUP/s]},
}}

\begin{tikzpicture}[scale=.85]
    \begin{axis}[ECM_back_style,at={(0,0)}]
        \addplot +[ybar,blue!60!white,mark=none] table [x = {N^3},y = {ECM LC Tnol}] {\datatwo};
        \addplot +[ybar,blue!30!white,mark=none] table [x = {N^3},y = {ECM LC Tl1l2}] {\datatwo};
        \addplot +[ybar,orange,mark=none] table [x = {N^3},y = {ECM LC Tl2l3}] {\datatwo};
        \addplot +[ybar,orange!50!white,mark=none] table [x = {N^3},y = {ECM LC Tl3mem}] {\datatwo};
    \end{axis}

    \begin{axis}[ECM_front_style,at={(0,0)}]
        \addplot+ [mark=x,black] table[x = {N^3},y = {Benchmark cycl}]{\datatwo};
        \addplot+ [mark=none,line width=.05cm,red] table[x = {N^3},y = {ECM LC Tol}]{\datatwo};
        \addplot+ [mark=o,blue] table[x = {N^3},y = {Roofline LC cycl}]{\datatwo};
    \end{axis}

    \begin{axis}[ECM_secondary_axis_style,at={(0,0)},
      every axis y label/.style={at={(current axis.east)},right=2cm,rotate=-90,font=\small,xshift=-1.5cm}]
    \end{axis}

    \path [-,thick,draw] (-1.1,4.75) -- (-.85,4.75);
      \node at (-.975 ,4.75) {\tiny$\times$};
      \node [anchor=west] at (-.95 ,4.75) {\tiny Benchmark};
    \path [-,thick,draw,blue] (.75,4.75) -- (1.05 ,4.75);
      \node [blue] at (.9,4.7375) {\large$\circ$};
      \node [anchor=west] at (.9,4.75) {\tiny Roofline};
    \path [-,thick,draw,red] (2.25 ,4.75) -- (2.45 ,4.75);
      \node [anchor=west] at (2.3,4.75) {\tiny $T_{\scaleto{\mathrm{comp}}{3pt}}$};
    \path [-,line width=.2cm,draw,blue!60!white] (3.3 ,4.75) -- (3.5 ,4.75);
      \node [anchor=west] at (3.35 ,4.75) {\tiny $T_{\scaleto{\mathrm{RegL1}}{3pt}}$};
    \path [-,line width=.2cm,draw,blue!30!white] (4.3 ,4.75) -- (4.5,4.75);
      \node [anchor=west] at (4.4,4.75) {\tiny $T_{\scaleto{\mathrm{L1L2}}{3pt}}$};
    \path [-,line width=.2cm,draw,orange] (5.4 ,4.75) -- (5.6,4.75);
      \node [anchor=west] at (5.5,4.75) {\tiny $T_{\scaleto{\mathrm{L2L3}}{3pt}}$};
    \path [-,line width=.2cm,draw,orange!50!white] (6.55 ,4.75) -- (6.75 ,4.75);
      \node [anchor=west] at (6.65 ,4.75) {\tiny $T_{\scaleto{\mathrm{L3Mem}}{3pt}}$};
\end{tikzpicture}
    \caption{\label{fig:3D3rheterostarconstdoubleSKYLC} ECM and Roofline predictions for SKX, based on layer conditions, with benchmark results for 3D/3r/heterogeneous/star/constant/double stencil.}
  \end{minipage}
\end{figure*}

\subsection{A Long-ranged Stencil on Skylake X (Intel Xeon Gold 6148)}\label{sec:example:longrange}


The second example showcases a long-ranged heterogeneous star stencil on the Skylake X architecture
(Intel Xeon Gold 6148), the stencil source code is shown in
Listing~\ref{lst:3D-3r-hetero-star-const-double}. It features more floating point operations
compared to the previous kernel because of the heterogeneous coefficients classification, but
also higher memory traffic due to the long stencil range (i.e., range = 3 or r3
classification). For brevity, only the stacked ECM prediction with layer conditions as cache
predictor is shown in Figure~\ref{fig:3D3rheterostarconstdoubleSKYLC}. All remaining information
and graphs can be found on the INSPECT website: \url{https://git.io/fjq2a}.

Qualitatively, both---Roofline and ECM prediction---represent the measured performance behavior
well. The Roofline model is a bit too optimistic and the ECM model a bit too pessimistic. The reason
for that is the new organization of the cache hierarchy in the Skylake microarchitecture, seen in
Figure~\ref{fig:architecture:skx} and discussed in Sec.~\ref{sec:architectures}. At the moment it
is not possible to correctly model the data transfers between L2 and L3 cache, in combination with
main memory. With the ECM model a worst case scenario is assumed, such that all data dropped or
evicted from L2 is passed onto L3. Taking that into account and with better knowledge of the actual
caching algorithms and heuristics, the ECM prediction would become faster and more closely match
the measured data.

The layer condition analysis, correlates very well with the measured data and all
relevant breaks (i.e., plateaus) can be seen. The slow performance in the beginning until $120^3$
is again related to the high $T_\mathrm{comp}$ fraction and scalar loads of the remainder
loop.

\begin{lstlisting}[language=C,label={lst:3D-3r-hetero-star-const-double},float,floatplacement=H,caption={3D stencil code with radius 3, constant hetero-geneous coefficients, star structure and double datatype.}]
double a[M][N][P], b[M][N][P];
double c0, c1, c2, c3, c4, c5, c6, c7, c8, c9;
double c10, c11, c12, c13, c14, c15, c16, c17, c18;

for(int k = 3; k < M-3; k++) {
  for(int j = 3; j < N-3; j++) {
    for(int i = 3; i < P-3; i++) {
      b[k][j][i] = c0 * a[k][j][i]
          + c1  * a[k][j][i-1] + c2  * a[k][j][i+1] + c3  * a[k-1][j][i]
          + c4  * a[k+1][j][i] + c5  * a[k][j-1][i] + c6  * a[k][j+1][i]
          + c7  * a[k][j][i-2] + c8  * a[k][j][i+2] + c9  * a[k-2][j][i]
          + c10 * a[k+2][j][i] + c11 * a[k][j-2][i] + c12 * a[k][j+2][i]
          + c13 * a[k][j][i-3] + c14 * a[k][j][i+3] + c15 * a[k-3][j][i]
          + c16 * a[k+3][j][i] + c17 * a[k][j-3][i] + c18 * a[k][j+3][i];
}}}
\end{lstlisting}

Data shown on the INSPECT website for full-socket thread scaling shows that the prediction fits
perfectly to the measured data. Also cache blocking for the L2-3D layer condition works very well,
due to the larger L2 cache in this architecture. In contrast to that, L3-3D cache blocking works
very poorly, as is in accordance with Intel's recommendations: ``Using just the last level
cache size per core may result in non-optimal use of available on-chip
cache''~\cite{intel_optimization_reference} (p. 41).

Overall it can be said, that except for the uncertainty with the L2-L3-MEM caching behavior, the
applied Skylake machine model works well and gives accurate predictions.

Performance optimization potential is again predicted by the plateaus (for spatial blocking)
and contributions (for temporal blocking). Spatial blocking to $N_\mathrm{block}^{1/3} < 300$ may
increase performance by up to 30\%. Temporal blocking would only make sense, in comparison to spatial,
if it is done in the L2 cache, stripping the two upper contributions off the non-overlapping ECM
prediction and possibly hitting the instruction throughput bottleneck ($T_\mathrm{comp}$) at 40\,cy/CL.

\subsection{Comparison of a Short-ranged Box Stencil on Broadwell and Skylake X}\label{sec:example:comparison}


\begin{figure*}[b]
    \hspace{-.75cm}
    \subcaptionbox{Broadwell (BDW)\label{fig:3D1rheteroboxconstdoubleBDWLC}}{

\pgfplotstableread[col sep=comma]{Plots/3D-1r-hetero-box-const-double-BDW.csv}\datatwo

\pgfplotsset{ECM_front_style/.style={
      width=8.25cm,
      height=6cm,
      xlabel={Grid Size (N$^{1/3}$)},
      ylabel={Inverse Throughput [cy/CL]},
      title style={yshift=.5cm},
      every axis y label/.style={at={(current axis.west)},left=1.25cm,rotate=90,font=\tiny},
      every axis x label/.style={at={(current axis.south)},yshift=-.75cm,font=\tiny},
      grid=major,
      ymin=0,ymax=139,
      ytick={0,20,...,120},
      ytick pos=left,
      xmin=10,xmax=1020,
      xtick={100,400,...,1000},
      mark repeat=10,
      tick label style={font=\footnotesize},
      label style={font=\small},
      ylabel style={anchor=north}
}}

\pgfplotsset{ECM_back_style/.style={
      width=8.25cm,
      height=6cm,
      xlabel={},
      ylabel={},
      every axis y label/.style={at={(current axis.west)},left=1.25cm,rotate=90,font=\tiny},
      every axis x label/.style={at={(current axis.south)},yshift=-.75cm,font=\tiny},
      ymin=0,ymax=139,
      ytick={0,20,...,120},
      ybar stacked,
      bar width=1.75pt,
      xmin=10,xmax=1020,
      xtick={100,400,...,1000},
      tick label style={font=\footnotesize,white},
      label style={font=\small},
      ylabel style={anchor=north}
}}

\pgfplotsset{ECM_secondary_axis_style/.style={
      ECM_front_style,
      title={},
      grid=none,
      axis y line*=right,
      axis x line=none,
      ymin=1, ymax=140,
      ytick={20,40,...,120},
      yticklabels={150,185,230,305,460,920},
      ytick pos=right,
      y dir=reverse,
      ylabel={Performance [MLUP/s]},
}}

\begin{tikzpicture}[scale=.85]
    \begin{axis}[ECM_back_style,at={(0,0)}]
        \addplot +[ybar,blue!60!white,mark=none] table [x = {N^3},y = {ECM LC Tnol}] {\datatwo};
        \addplot +[ybar,blue!30!white,mark=none] table [x = {N^3},y = {ECM LC Tl1l2}] {\datatwo};
        \addplot +[ybar,orange,mark=none] table [x = {N^3},y = {ECM LC Tl2l3}] {\datatwo};
        \addplot +[ybar,orange!50!white,mark=none] table [x = {N^3},y = {ECM LC Tl3mem}] {\datatwo};
    \end{axis}

    \begin{axis}[ECM_front_style,at={(0,0)}]
        \addplot+ [mark=x,black] table[x = {N^3},y = {Benchmark cycl}]{\datatwo};
        \addplot+ [mark=none,line width=.05cm,red] table[x = {N^3},y = {ECM LC Tol}]{\datatwo};
        \addplot+ [mark=o,blue] table[x = {N^3},y = {Roofline LC cycl}]{\datatwo};
    \end{axis}

    \begin{axis}[ECM_secondary_axis_style,at={(0,0)},
      every axis y label/.style={at={(current axis.east)},right=2cm,rotate=-90,font=\small,xshift=-1.5cm}]
    \end{axis}

    \path [-,thick,draw] (-1.1,4.75) -- (-.85,4.75);
      \node at (-.975 ,4.75) {\tiny$\times$};
      \node [anchor=west] at (-.95 ,4.75) {\tiny Benchmark};
    \path [-,thick,draw,blue] (.75,4.75) -- (1.05 ,4.75);
      \node [blue] at (.9,4.7375) {\large$\circ$};
      \node [anchor=west] at (.9,4.75) {\tiny Roofline};
    \path [-,thick,draw,red] (2.25 ,4.75) -- (2.45 ,4.75);
      \node [anchor=west] at (2.3,4.75) {\tiny $T_{\scaleto{\mathrm{comp}}{3pt}}$};
    \path [-,line width=.2cm,draw,blue!60!white] (3.3 ,4.75) -- (3.5 ,4.75);
      \node [anchor=west] at (3.35 ,4.75) {\tiny $T_{\scaleto{\mathrm{RegL1}}{3pt}}$};
    \path [-,line width=.2cm,draw,blue!30!white] (4.3 ,4.75) -- (4.5,4.75);
      \node [anchor=west] at (4.4,4.75) {\tiny $T_{\scaleto{\mathrm{L1L2}}{3pt}}$};
    \path [-,line width=.2cm,draw,orange] (5.4 ,4.75) -- (5.6,4.75);
      \node [anchor=west] at (5.5,4.75) {\tiny $T_{\scaleto{\mathrm{L2L3}}{3pt}}$};
    \path [-,line width=.2cm,draw,orange!50!white] (6.55 ,4.75) -- (6.75 ,4.75);
      \node [anchor=west] at (6.65 ,4.75) {\tiny $T_{\scaleto{\mathrm{L3Mem}}{3pt}}$};
\end{tikzpicture}
    }
    \hspace{-.75cm}
    \subcaptionbox{Skylake X (SKX)\label{fig:3D1rheteroboxconstdoubleSKYLC}}{

\pgfplotstableread[col sep=comma]{Plots/3D-1r-hetero-box-const-double-SKX_SNC.csv}\datatwo

\pgfplotsset{ECM_front_style/.style={
      width=8.25cm,
      height=6cm,
      xlabel={Grid Size (N$^{1/3}$)},
      ylabel={Inverse Throughput [cy/CL]},
      title style={yshift=.5cm},
      every axis y label/.style={at={(current axis.west)},left=1.25cm,rotate=90,font=\tiny},
      every axis x label/.style={at={(current axis.south)},yshift=-.75cm,font=\tiny},
      grid=major,
      ymin=0,ymax=139,
      ytick={0,20,...,120},
      ytick pos=left,
      xmin=10,xmax=1020,
      xtick={100,400,...,1000},
      mark repeat=10,
      tick label style={font=\footnotesize},
      label style={font=\small},
      ylabel style={anchor=north}
}}

\pgfplotsset{ECM_back_style/.style={
      width=8.25cm,
      height=6cm,
      xlabel={},
      ylabel={},
      every axis y label/.style={at={(current axis.west)},left=1.25cm,rotate=90,font=\tiny},
      every axis x label/.style={at={(current axis.south)},yshift=-.75cm,font=\tiny},
      ymin=0,ymax=139,
      ytick={0,20,...,120},
      ybar stacked,
      bar width=1.75pt,
      xmin=10,xmax=1020,
      xtick={100,400,...,1000},
      tick label style={font=\footnotesize,white},
      label style={font=\small},
      ylabel style={anchor=north}
}}

\pgfplotsset{ECM_secondary_axis_style/.style={
      ECM_front_style,
      title={},
      grid=none,
      axis y line*=right,
      axis x line=none,
      ymin=1, ymax=140,
      ytick={20,40,...,120},
      yticklabels={160,190,240,320,480,960},
      ytick pos=right,
      y dir=reverse,
      ylabel={Performance [MLUP/s]},
}}

\begin{tikzpicture}[scale=.85]
    \begin{axis}[ECM_back_style,at={(0,0)}]
        \addplot +[ybar,blue!60!white,mark=none] table [x = {N^3},y = {ECM LC Tnol}] {\datatwo};
        \addplot +[ybar,blue!30!white,mark=none] table [x = {N^3},y = {ECM LC Tl1l2}] {\datatwo};
        \addplot +[ybar,orange,mark=none] table [x = {N^3},y = {ECM LC Tl2l3}] {\datatwo};
        \addplot +[ybar,orange!50!white,mark=none] table [x = {N^3},y = {ECM LC Tl3mem}] {\datatwo};
    \end{axis}

    \begin{axis}[ECM_front_style,at={(0,0)}]
        \addplot+ [mark=x,black] table[x = {N^3},y = {Benchmark cycl}]{\datatwo};
        \addplot+ [mark=none,line width=.05cm,red] table[x = {N^3},y = {ECM LC Tol}]{\datatwo};
        \addplot+ [mark=o,blue] table[x = {N^3},y = {Roofline LC cycl}]{\datatwo};
    \end{axis}

    \begin{axis}[ECM_secondary_axis_style,at={(0,0)},
      every axis y label/.style={at={(current axis.east)},right=2cm,rotate=-90,font=\small,xshift=-1.5cm}]
    \end{axis}

    \path [-,thick,draw] (-1.1,4.75) -- (-.85,4.75);
      \node at (-.975 ,4.75) {\tiny$\times$};
      \node [anchor=west] at (-.95 ,4.75) {\tiny Benchmark};
    \path [-,thick,draw,blue] (.75,4.75) -- (1.05 ,4.75);
      \node [blue] at (.9,4.7375) {\large$\circ$};
      \node [anchor=west] at (.9,4.75) {\tiny Roofline};
    \path [-,thick,draw,red] (2.25 ,4.75) -- (2.45 ,4.75);
      \node [anchor=west] at (2.3,4.75) {\tiny $T_{\scaleto{\mathrm{comp}}{3pt}}$};
    \path [-,line width=.2cm,draw,blue!60!white] (3.3 ,4.75) -- (3.5 ,4.75);
      \node [anchor=west] at (3.35 ,4.75) {\tiny $T_{\scaleto{\mathrm{RegL1}}{3pt}}$};
    \path [-,line width=.2cm,draw,blue!30!white] (4.3 ,4.75) -- (4.5,4.75);
      \node [anchor=west] at (4.4,4.75) {\tiny $T_{\scaleto{\mathrm{L1L2}}{3pt}}$};
    \path [-,line width=.2cm,draw,orange] (5.4 ,4.75) -- (5.6,4.75);
      \node [anchor=west] at (5.5,4.75) {\tiny $T_{\scaleto{\mathrm{L2L3}}{3pt}}$};
    \path [-,line width=.2cm,draw,orange!50!white] (6.55 ,4.75) -- (6.75 ,4.75);
      \node [anchor=west] at (6.65 ,4.75) {\tiny $T_{\scaleto{\mathrm{L3Mem}}{3pt}}$};
\end{tikzpicture}
    }
    \caption{ECM and Roofline model predictions based on layer conditions, with benchmark results for 3D/1r/heterogenous/box/constant/double stencil on Broadwell and Skylake X architectures.}
\end{figure*}
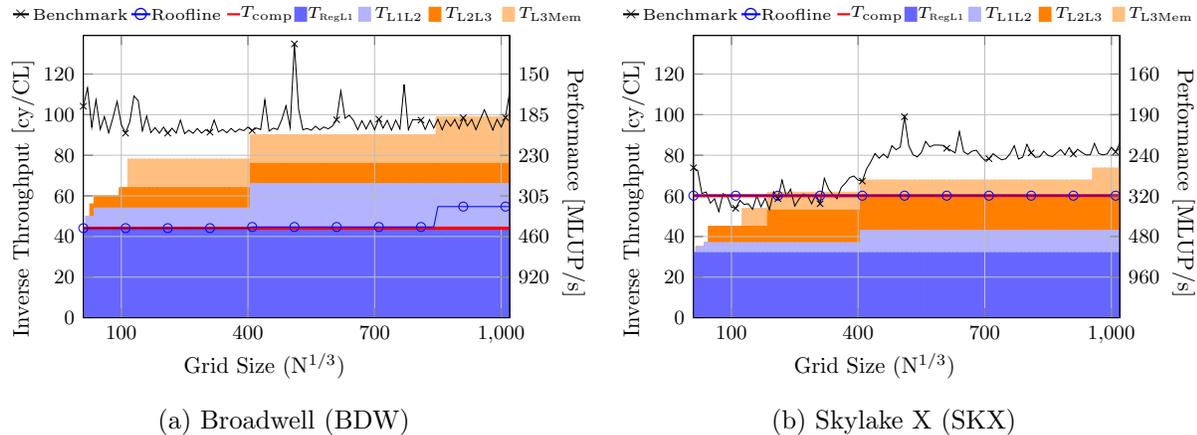

\begin{lstlisting}[language=C,label={lst:3D-1r-hetero-box-const-double},float,floatplacement=!b,caption={3D stencil code with radius 1, variable hetero-geneous coefficients, box structure and double datatype.}]
double a[M][N][P], b[M][N][P], W[27][M][N][P];

for(int k = 1; k < M - 1; ++k) {
  for(int j = 1; j < N - 1; ++j) {
    for(int i = 1; i < P - 1; ++i) {
      b[k][j][i] = W[0][k][j][i] * a[k][j][i]
          + W[1][k][j][i]  * a[k-1][j-1][i-1] + W[2][k][j][i]  * a[k][j-1][i-1]
          + W[3][k][j][i]  * a[k+1][j-1][i-1] + W[4][k][j][i]  * a[k-1][j][i-1]
          + W[5][k][j][i]  * a[k][j][i-1]     + W[6][k][j][i]  * a[k+1][j][i-1]
          + W[7][k][j][i]  * a[k-1][j+1][i-1] + W[8][k][j][i]  * a[k][j+1][i-1]
          + W[9][k][j][i]  * a[k+1][j+1][i-1] + W[10][k][j][i] * a[k-1][j-1][i]
          + W[11][k][j][i] * a[k][j-1][i]     + W[12][k][j][i] * a[k+1][j-1][i]
          + W[13][k][j][i] * a[k-1][j][i]     + W[14][k][j][i] * a[k+1][j][i]
          + W[15][k][j][i] * a[k-1][j+1][i]   + W[16][k][j][i] * a[k][j+1][i]
          + W[17][k][j][i] * a[k+1][j+1][i]   + W[18][k][j][i] * a[k-1][j-1][i+1]
          + W[19][k][j][i] * a[k][j-1][i+1]   + W[20][k][j][i] * a[k+1][j-1][i+1]
          + W[21][k][j][i] * a[k-1][j][i+1]   + W[22][k][j][i] * a[k][j][i+1]
          + W[23][k][j][i] * a[k+1][j][i+1]   + W[24][k][j][i] * a[k-1][j+1][i+1]
          + W[25][k][j][i] * a[k][j+1][i+1]   + W[26][k][j][i] * a[k+1][j+1][i+1];
}}}
\end{lstlisting}

Finally, we present a comparison of Broadwell (Intel Xeon E5-2697v4) and Skylake X
(Intel Xeon Gold 6148) with a short-ranged box stencil with heterogeneous constant coefficients,
cf. Listing~\ref{lst:3D-1r-hetero-box-const-double}. Compared to star stencils, box stencils
need more loads and registers, which may have a large performance impact. In Figures~\ref{fig:3D1rheteroboxconstdoubleBDWLC}
and~\ref{fig:3D1rheteroboxconstdoubleSKYLC} benchmark data and model predictions are shown. On the
INSPECT website a complete list of graphs, data sets and modeling information may be viewed:
\url{https://git.io/fjqav} for Broadwell and \url{https://git.io/fjqaU} for Skylake X.

On Broadwell performance does not seem to be impacted by data traffic nor in-core execution
($T_\mathrm{comp}$). Reasons for the poor and almost constant performance across all grid
sizes is a register dependency chain that is visible in the assembly code, to be seen on the
INSPECT website under ``Kernel Source Code'' by clicking on the ``Assembly Code'' button, but
undetected by IACA. This dependency chain slows down the execution so much, that all other effects
are suppressed and the performance becomes independent of the grid size. Since the number of
available registers has been doubled with the introduction of the Skylake X architecture, this
disastrous effect is eliminated there. Instead, until $300^3$ the in-core bottleneck
$T_\mathrm{comp}$ dominates and limits the reciprocal throughput to $\sim60$\,cy/CL.
This prediction, originating from an IACA analysis, is obviously too pessimistic, since
measurements show better performance compared to ECM and Roofline models. Looking into the IACA
analysis, it only gives an explanation for 46 out of the 60 cycles and adds 14 cycles
based on an unknown heuristic. Considering this, 46 cycles per cacheline would explain the
measured performance much better and calls for a better in-core model, as aimed for by the OSACA
project~\cite{OSACA}. Beyond $N\approx400^3$, data transfers become more dominant and slow down
the execution, as qualitatively predicted by the ECM model. Roofline sticks with the 60\,cy/CL,
because no single memory level surpasses them. In light of the discrepancy between modeled and
measured performance, the graph can not be used to guide performance optimization, but it sheds
light on the IACA misprediction at hand.

For Skylake, simple spatial blocking with $N_\mathrm{block}^{1/3} < 400$ is advisable. Temporal
blocking would not yield better results, because of the hard $T_\mathrm{comp}$ limit.

\section{How to Make Use of INSPECT}\label{sec:howto}

When developing stencil-driven applications and especially when publishing performance results
based on stencil codes, authors have to compare to a suitable, well-understood baseline.
In order to make use of INSPECT in this context, the user must first classify their
stencil according to our scheme and select a microarchitecture and CPU model
from the INSPECT website that is similar or identical to their own. If that is not possible,
INSPECT provides the toolchain and automation to compile a new baseline for future reference.

Depending on the programming language and software architecture, stencil patterns in applications
may be hidden under several abstraction layers but come to light during detailed performance
analysis. It is also the user's task to isolate the stencil code in order to be able to measure its
performance. This may be done either ``in situ'' via suitable instrumentation or by writing a proxy
application that only performs stencil updates.
Once a stencil is classified and a comparison is established, optimization strategies may be guided
by the INSPECT ECM model report: Spatial blocking should bring the performance of large data sets
on the level of smaller data set sizes by better use of caches (moving to a plateau left in the
plots), whereas temporal blocking strategies eliminate data transfers to lower memory hierarchy
levels (peeling off layers in the stacked ECM plot contributions).

If the measured stencil performance in the application code does not coincide with the INSPECT data
and model at least qualitatively, as seen in Sec.~{\ref{sec:example:comparison}}, the culprit is
usually the compiler not generating efficient code, but other scenarios are possible: specific
hardware features in the user's benchmarking setup (e.g., different DIMM organization), unfavorable
system settings (e.g., small OS pages, uncontrolled clock speed, Cluster-on-Die settings, disabled
prefetchers), simple benchmarking mistakes such as wrong or no thread affinity, etc.. Whatever
the reason, it will be worth investigating, which usually leads to better insight.
%

\section{Related Work}\label{sec:related}

Our work comprises three parts: stencil classification and generation with STEMPEL,
benchmarking and modeling with Kerncraft, and presentation and publication of results on the INSPECT
website.

Collecting and presenting benchmark results is a common approach for a variety of reasons. To name a few examples:
\begin{itemize}
    \item The TOP500~\cite{top500} ranks the performance of HPC systems world-wide based on the High Performance LINPACK~\cite{linpack} benchmark performance.
    \item The HPCG benchmark~\cite{HPCG} takes the same approach as the TOP500, with a different benchmark.
    \item SPEC~\cite{SPEC} has a spectrum of benchmarks suites for different aspects and allows its
      members to publish the results on their website. Their suites come with real applications
      embedded as test cases. They produce detailed reports on the runtime environment, with the
      goal of comparing the performance of systems.
    \item The STREAM benchmark~\cite{McCalpin1995} is the de facto standard for measuring
      main memory bandwidth. The website has results for machines in tabulated form.
    \item The HPC Challenge Benchmark Suite~\cite{hpcc} combines multiple benchmarks and allows
      users to publish results through HPCC's website.
\end{itemize}
All of these benchmark collections are focused on comparing machine performance by a set of
predefined benchmarks, which is extremely valuable for purchasing decisions and as a reference for
researchers and developers. In contrast, we try to explain the observed performance based on the
characteristics of the stencil structure, which is usually defined by the underlying model and
discretization. This makes it more informative and adaptable for a particular developer to compare
and explain their own code's performance with similarly structured reference implementations
provided by our framework.

The Ginkgo Performance Explorer\cite{ginkgo_pe} focuses on presenting performance data gathered by
automatic benchmarks as part of a continuous integration (CI) pipeline. This project is generically
applicable to other workflows, but lacks the focus on a specific field to allow the fine grained
presentation of model predictions and measurements as is done by INSPECT, nor does it comprise any
modeling component.

Methodologies for performance analysis most often fall into the category of performance models,
such as the already mentioned ECM and Roofline models. Their application to specific stencils or
stencil-based algorithms was at the focus of intense research~\cite{Sch_fer_2013, malas_2015,
Ghysels_2015, datta09, Hornich_2018}. Our concept goes beyond these approaches in that it enables
easy reproduction of performance numbers and encourages discussion via an open workflow.

Datta et al.\ published a study of a stencil kernel on multiple architectures
in 2009~\cite{datta09}. It is based on the same modeling principles but does not provide a unified
process and presentation reusable for other kernels.

\section{Conclusion and Outlook}\label{sec:conc}

We have presented a comprehensive code generation, performance evaluation, modeling, and data
presentation framework for stencil algorithms. It includes a classification scheme for stencil
characteristics, a data collection methodology, automatic analytic performance modeling via advanced
tools and a publicly available website that allows to browse results and models across a variety of
processor architectures in a convenient way.

The presented baseline performance and model data provides valuable insight and points of comparison for
developers who have to write performance critical stencil code. The automatic spatially optimized
version is given as an optimization example. INSPECT already contains a large range of
different stencil parameters and will be continuously extended to eventually comprise a full
coverage of the parameter space. To this end, we plan on optimizing the tool chain to reduce the
total runtime considerably. The choice of an analytic performance model over machine learning was deliberate as not only prediction but also insight into bottlenecks is desired.

Kerncraft support of non-Intel architectures is still rudimentary. Support for AMD's latest x86
implementations is already available and we have presented its preliminary use, while ARM will
require more effort but is on our shortlist of upcoming features. These additions will be
integrated into future updates of the INSPECT website.

An interesting spin-off of this work would be the integration of more web-enabled tools, such as the
layer-condition analysis~\cite{layercondition}, into INSPECT to allow users to interactively analyze
their own code. Compiler explorer~\cite{godbolt} would be one potential tool to inspect
compiler behavior for different architectures.

A generalization from stencils to dense linear algebra and streaming kernels is straightforward
from Kerncraft's perspective, but the classification scheme would have to be extended.

\section*{Acknowledgements}
The authors gratefully acknowledge funding of the German Federal Ministry for Education and Research
(BMBF) via the SeASiTe, SKAMPY, and METACCA projects. We are thankful
for support by the Bavarian Competence Network for Scientific High Performance Computing
(KONWIHR).

\openaccess

\bibliography{bibfile}
\end{document}